# Dominant 1/3-filling Correlated Insulator States and Orbital Geometric Frustration in Twisted Bilayer Graphene

Haidong Tian[1]*, Emilio Codecido[1]*, Dan Mao[2]*, Kevin Zhang[2], Shi Che[1], Kenji Watanabe[3], Takashi Taniguchi[4], Dmitry Smirnov[5], Eun-Ah Kim[2, 6, 7, 8†], Marc Bockrath[1‡], Chun Ning Lau[1§]

[1] Department of Physics, The Ohio State University, Columbus, OH 43221.
[2] Laboratory of Atomic and Solid State Physics, Cornell University, 142 Sciences Drive, Ithaca NY 14853-2501, USA
[3] Research Center for Electronic and Optical Materials, National Institute for Materials Science, 1-1 Namiki, Tsukuba 305-0044, Japan
[4] Research Center for Materials Nanoarchitectonics, National Institute for Materials Science, 1-1 Namiki, Tsukuba 305-0044, Japan
[5] National High Magnetic Field Laboratory, Tallahassee, FL 32310
[6] Radcliffe Institute for Advanced Study at Harvard, Harvard University, 10 Garden Street, Cambridge MA 02138, USA
[7] Department of Physics, Harvard University, 17 Oxford Street, Cambridge MA 02138, USA
[8] Department of Physics, Ewha Womans University

* These authors contributed equally to the manuscript.

**Geometric frustration is a phenomenon in a lattice system where not all interactions can be satisfied, the simplest example being antiferromagnetically coupled spins on a triangular lattice. Frustrated systems are characterized by their many nearly degenerate ground states, leading to non-trivial phases such as spin ice and spin liquids. To date most studies are on geometric frustration of spins; much less explored is orbital geometric frustration. For electrons in twisted bilayer graphene (tBLG) at denominator 3 fractional filling, Coulomb interactions and the Wannier orbital shapes are predicted to strongly constrain spatial charge ordering, leading to geometrically frustrated ground states that produce a new class of correlated insulators (CIs). Here we report the observation of dominant denominator 3 fractional filling insulating states in large angle tBLG; these states persist in magnetic fields and display magnetic ordering signatures and tripled unit cell reconstruction. These results are in agreement with a strong-coupling theory of symmetry-breaking of geometrically frustrated fractional states.**

Correlated insulating states observed in tBLG and moiré transition metal dichalcogenides (TMDs) have been intensively studied experimentally and theoretically[1-10], resulting from an interplay between the periodic superlattice, the magnetic field, and various interactions. These states can be classified into different classes of insulators via the equation

$$(n/n_0) = tb + \bar{n} \quad (1)$$

where $n$ is the charge density, $n_0$ is the density corresponding to one charge per unit cell, $b = BAe/h$ is the number of flux quanta per unit cell, $A$ is the area of the superlattice unit cell, and $t$ and $\bar{n}$ are constants. Here $t$ is the Chern number, which determines the transverse conductivity $\sigma_{xy} = te^2/h$ [11], while the additional quantum number $\bar{n}$ can be viewed as the number of bound charges per unit cell [12]. Both $\bar{n}$ and $t$ can take on either integer or fractional values, indicating the presence of physically distinct ground states[13][14].

† Email: ek436@cornell.edu
‡ Email: bockrath.31@osu.edu
§ Email: lau.232@osu.edu

To date the vast majority of insulating states in tBLG are observed at integer $t$ and $\bar{n}$, i.e. corresponding to integer quantum Hall (QH) or Chern insulator states. States with fractional $\bar{n}$ values have been observed in moire TMDs [4-6], and are attributed to generalized Wigner crystal states where electrons are localized in a fraction of the moiré unit cells. Very recently, signatures of states with fractional $\bar{n}$ values have been observed in a magic angle (MA) tBLG at twist angle $\vartheta=1.06°$[15], in twisted double-bilayer graphene[16, 17], and in twisted trilayer graphene[18], but are much weaker than those at integer integer $\bar{n}$. These states are attributed to band structure-driven effects that originate in momentum-space[15] [18], though fine tuning is required. More generally, for systems at or close to the magic angle, the close competition between the charge ordered, symmetry-broken, and fractional Chern insulators in the systems challenges sharp theoretical understanding of the charge ordered insulating states.

To date, very little consideration was given to the effect of geometric frustration of charge carriers in twisted graphene systems with triangular moiré lattices. Here we report the observation of robust Chern-zero fractional correlated insulating states at a number of 1/3-fillings in tBLG with larger-than magic twist angles ($\vartheta=1.32 – 1.59°$). These states strongly dominate over those at integer fillings, and persist in perpendicular and parallel magnetic fields, and are accounted for by an orbital geometric frustration model, which is based on the 3-leaved shape of the Wannier orbitals while considering Coulomb interactions, the site occupancy, valley polarization, and exchange interactions between the charge-ordered lattice sites. In a device with a relatively small angle (but still larger than the MA), we observe behavior consistent with a predicted "brick-wall" charge-ordered state at 1/3 fractional fillings[8, 19]. Upon applying a parallel magnetic field $B_{\parallel}$, we find that the $\bar{n}=\pm 8/3$ states exhibit ferromagnetic ordering, while the $\bar{n}=\pm 4/3$ states exhibit antiferromagnetic ordering, with a transition to a spin-aligned state observable at finite field. Moreover, in a device with a larger twist angle and expected stronger second nearest neighbor interactions, we observe a tripled unit cell, consistent with a predicted "armchair" phase[8, 19]. These 1/3 fractional states therefore originate from the Wannier orbital shape and resulting geometric frustration produced by Coulomb interactions, with the phase tunable by doping and twist angle. This shows that a significant class of phenomena in tBLG can be understood via a real space model.

Figure 1 presents data from device D1 with $\vartheta=1.32°$. Figure 1a plots its two-terminal conductance $G$ versus back gate voltage $V_{bg}$ (bottom axis) and $\bar{n}$ (top axis). At $T=0.4$ K (blue curve), conductance valleys are observed at the global charge neutrality point (CNP) of the moiré miniband; at half filling $\bar{n}=2$, a small conductance dip is observed at $\bar{n}=+2$ and a kink at $\bar{n}=-2$. These observations are similar to those observed in MA-tBLG. In contrast to MA-tBLG, however, the device exhibits many fractional filling features: deep conductance valleys are also observed at $\bar{n}=\pm 8/3$ and $\pm 4/3$, and shallower dips or shoulders emerge at $\bar{n}=\pm 2/3$ and $10/3$. Remarkably, in this device with $\vartheta$ larger than the MA, these 1/3-filling features are much stronger than the integer fillings features; their conductivity, 0.1-10 µS, are much smaller than the conductance quantum $e^2/h\sim 39$ µS, underscoring their insulating nature.

To further investigate these fractional fillings features in more detail, we measure their $T$-dependence. Figure 1a shows that the conductance valleys rise and broaden with $T$ until they merge into a broad background that increases at high temperatures. An Arrhenius plot of $G$ in the conductance valleys at $\bar{n}=\pm 8/3$ shows that they can be fit to a double exponential function: a large gap $\Delta_1$ ~21.0 (16.5) meV and a small one $\Delta_2$ ~2.6 (1.4) meV are extracted for the $\bar{n}$ =+8/3 ($\bar{n}$ =-8/3) state (Fig. 1b). We attribute the larger gap, which is responsible to the high temperature behavior, to that of the insulating state at $\bar{n}$=±8/3; the smaller gap, on the other hand, likely arises from competing mechanisms at low temperatures from, e.g. variable range hopping through localized states (Fig. 1b). On the other hand, the $\bar{n}$=±4/3 valleys behave qualitatively differently: as $T$ cools below 70K (Fig. 1c right axis), the conductance first slowly increases, but begins to drop sharply when the temperature reaches below ~10K; fitting the low temperature regime to a thermal activation model reveals a small gap of ~10K, which also agrees with the onset temperature of the conductance's non-monotonic dependence. Thus, the $\bar{n}$=±4/3 states also appear to host a significantly smaller gap of ~1 meV, where the conductance appears activated, with saturation due to variable range hopping conduction at lowest temperature.

To understand the emergence of these fractional gapped states, we start from ref. [8], which predicts a Chern-zero insulating state in MA-tBLG at 1/3 fractional fillings [20-22]. Unlike the momentum-space picture adopted in [15] and [18], this strong coupling real-space picture does not readily capture the band-structure sensitive Chern-insulator physics; rather, the key insight lies with the three-leaved geometry of the WOs[20-22] that quite generally leads to the robust presence of incompressible states at 1/3 fractional fillings and emergent geometric frustration[20-22]. We generalize the model of ref. [8] to $\bar{n}$=±8/3 and $\bar{n}$=±4/3 for larger-than-magic tBLG, by constructing the WO's of each valley using the projection method [23], and including direct density-density interaction $V$'s and the direct exchange interaction $J$'s up to third nearest neighbor (see Fig. 2b and Supplementary Information (SI)). Notably, significant differences exist between the WOs constructed from the two valleys on the same site (Fig. 2a); thus the SU(4) spin-valley symmetry, which is conventionally assumed in calculations, is broken. Instead, the resulting Hamiltonian has a SU(2)× SU(2) ×U(1)$_v$ symmetry, that accounts for separate spin-rotation symmetry for each valley and valley polarization conservation.

Similar to the case at $\bar{n}$=-4+1/3 discussed in [8, 19], exchange interactions promote ferromagnetic spin and valley alignment, and the distinct valley WOs imply that Coulomb interactions favor valley-locking at AB/BA sublattice sites. On the other hand, charge sharing at higher fillings at $\bar{n}$=±8/3 or ±4/3 (counting from the empty band at $\bar{n}$ =-4) forces WO touching, resulting in close competition between a rich set of possibilities [19], a signature of a geometrically frustrated system; the balance between different interaction strengths determines the ground state.

At $\bar{n}$=±8/3, four electrons or holes are shared among three moiré cells, thus each moiré site is either empty or singly occupied. We find four competing charge configurations (Fig. 2c): two with a 3-fold unit cell enlargement (the "star" and "armchair" phases), and two with a 6-fold enlargement ("brick-wall" and "zigzag" phases). At $\vartheta$=1.32°, the brick wall phase is narrowly favored over all the other phases, though the enlarged unit cell is not expected to be detected in transport due to the residual local degeneracy[19] (see SI). Moreover, the direct exchange interactions favor aligning spin and valley between neighboring sites, thus we anticipate the $\bar{n}$=±8/3 ground state to be a spin and valley polarized ferromagnetic insulator. Thermally excited conduction in this state requires sites to be doubly occupied, thus the thermal activation gap on the order of the energy cost of double occupation, which we estimate is ~20meV (see SI), consistent with the experimentally observed gap size.

For $\bar{n}$=±4/3, eight electrons (holes) must be accommodated among three moiré cells, so at least two sites must be doubly occupied, leading to additional geometric frustration and richness. Direct Coulomb interactions favor the doubly occupied sites to form a valley polarized state, while Pauli exclusion requires such doubly occupied sites to be in a spin-singlet state (Fig. 2d) at zero field. At low temperatures, the conductance comes from the electrons on the doubly occupied sites flowing on top of the singly occupied background; the thermal activation gap is thus on the order of further neighbor interactions instead of on-site repulsion, leading to a smaller gap than at $\bar{n}=\pm8/3$, in agreement with observations. Moreover, our model suggests that the $\bar{n}=\pm8/3$ and $\bar{n}=\pm4/3$ states should respond differently to in-plane magnetic fields: a stabilized ferromagnetic phase and a singlet-triplet transition are expected, respectively.

We now investigate these 1/3-filling states in a magnetic field to probe the Landau level structure and spin states. Fig. 3a-b displays the conductance $G$ versus charge density and perpendicular magnetic field $B$, and the corresponding Wannier diagram identifying features in $G$ with the numbers indicating the Chern number $t$ in Eq. (1). At the CNP and $B=0$, symmetric Landau fans "radiate" into both electron- and hole- regimes, similar to those observed in MA-tBLG devices[1, 24-27]. The conductance of all fractional filling factor states decreases with $B$, as shown in Fig. 3c-d. (An exception is the state at $\bar{n}=+4/3$, where $G(B)$ is almost constant; this asymmetry from that at $\bar{n}=-4/3$ may arise from electron-hole asymmetry in band structure, or different effects of nearby Landau levels emanating from $\bar{n}=0$). While no Landau fan is observed at $\bar{n}=\pm8/3$, Landau fans emerge at $\bar{n}=\pm4/3$ and 2/3, appearing to the left (right) of the hole- (electron-) doped regime, respectively, with a two-fold degeneracy.

These asymmetric or one-sided Landau fans are reminiscent of those at $\bar{n}=\pm2$ and $\pm3$ in MA-tBLG devices, in which a phase transition and a reconstruction of the Fermi surface occurs, yielding a reset of the band structure to the Dirac point[24, 28, 29]. Here we see a similar reset, and the degeneracy of two indicates an isospin polarization. On the other hand, since the reset occurs at fractional filling factors in this large-angle device, rather than for the integer $\bar{n}$ values in magic-angle devices, both the underlying phase transition and the reconstructed ground state are likely to be very different.

The presence or absence of Landau fan at different fractional fillings indicate that they host different electronic ground states. To further explore their nature and to study magnetic ordering at $\bar{n}=\pm8/3$, we apply an in-plane magnetic field $B_{||}$, which induces Zeeman splitting. Fig. 3e-f plots $G(\bar{n}, B_{||})$ and line traces at $B_{||}=0T$, 3T, 6T, 12T and 18T, respectively. The conductance of the states at $\bar{n}=\pm8/3$ decreases with increasing in-plane field. This suggests an increasing gap, such as that of a spin-polarized state (Fig. 3g). This observation provides evidence for the predicted brick-wall state's valley polarized, spin ferromagnetic, fractional insulating ground state: here every site is covered by the extended WO's with spin and valley polarization, so spin flips are needed for conduction. Hence increasing $B_{||}$ raises the energy cost of spin flips and leads to enhanced activation gap or lower conductivity, as observed.

The situation at $\bar{n}=\pm4/3$ differs, however: conductance of the state at $\bar{n}=4/3$ increases with increasing $B_{||}$, suggesting closing of an energy gap, such as that of an antiferromagnetic state (Fig. 3h). Interestingly, the state at $\bar{n}=-4/3$ exhibits non-monotonic dependence on $B_{||}$, thus hinting at the closing and opening of a gap at $B_{||} \sim 8T$. Taken together, these varying behaviors of states at $\bar{n}=\pm4/3$ suggest the presence of states with several competing magnetic orders arising from orbital and spin geometric frustration, which are consistent with the many competing ground states of the "parent" integer states at $\bar{n}=\pm1$ [30-35]. The observed non-monotonic $G$ vs. $B_{||}$ dependence thus arises from competition between the spin singlet and spin-polarized states at the doubly occupied sites of $\bar{n}=\pm4/3$: $B_{||}$ introduces a Zeeman term that favors a spin-polarized, valley-anti-aligned state (Fig. 2d), but a finite $B_{||}$ is required for the Zeeman energy to overcome the Coulomb-driven valley polarization energy. Hence, increasing $B_{||}$ initially reduces the transport gap and increases conductivity, until the spin-triplet phase becomes the ground state, whence the increase in $B_{||}$ increases the transport gap, resulting in decreasing $G$. The observed non-monotonic behavior for the 4/3 filling state under in-plane magnetic field supports this expectation of frustration-induced competition between spin singlet and spin-polarized states at the doubly occupied sites of 4/3 filling. The maximum $G$ at $B_\parallel \sim 10$ T suggests an energy scale ~1 meV for the valley polarization energy scale.

Returning to the competing states in Fig. 2b, a different ground state can be realized by changing the balance between the various extended interaction terms. This can be achieved by tuning the twist angle – a large $\vartheta$ results not only in a smaller moiré unit but also modifications in the competition between the second and third nearest neighbor interactions. At a larger twist angle, our model predicts that the armchair phase becomes narrowly favored over the brick wall phase due to increasing of second nearest neighbor interaction with respect to the third nearest neighbor ones, resulting in a tripling of the unit cell relative to the moiré unit cell (see SI for detailed discussions).

These predictions, in particular the unit cell tripling, are borne out experimentally in a second device D2 that has a larger twist angle $\vartheta=1.59\pm0.02°$. Similar to D1, it displays prominent four-probe longitudinal resistance $R_{xx}$ features at $\bar{n}=\pm 8/3$ and $\pm 4/3$ (their thermal activation behavior is not well-resolved, due to the reduced correlation among charge carriers), while showing strongly suppressed peaks at $\bar{n}=\pm 2$, as shown in Fig. 4a. At finite magnetic field, multiple Landau fan features at 1/3-fractional fillings appear. Here the Landau fans emanate from their zero-field peaks in both doping directions (Fig. 4b). The resolved Landau levels (LLs) with corresponding quantum $t$ numbers are sketched in Fig. 4c. The Landau fan emanating from the CNP is electron-hole symmetric, where the degeneracy of the lowest LL is fully resolved, and the rest of the Landau fan is 4-fold degenerate, with spectra gaps observed at $t=4N$, where $N$ is an integer. Similar Landau fans with $t=4N$ spectral gaps are observed to emanate from $\bar{n}=-4/3$, where the degeneracy is doubled compared to D1 and is likely due to the broadening of the Hofstadter bands in D2 since the kinetic energy is larger for a larger twist angle. Interestingly, for the Landau fan centered at $\bar{n}=\pm 8/3$, spectral gaps are electron-hole asymmetric: states at $t=2, 6, 10...$ are resolved in the electron and hole side for $\bar{n}=8/3$ and $-8/3$, respectively, while those at $t=4$ and 8 are resolved on the other side. This asymmetry in LL gaps indicates that when tuning across $\bar{n}=\pm 8/3$, the Berry phase experienced by the charge carriers undergoing cyclotron orbits changes abruptly from 0 to $\pi$ (modulo $2\pi$), which in turn suggests a novel reset of the effective band structure. Note also that similar to D1, the trajectory of these gapped states is vertical, which implies zero Chern number.

Detailed examination of the Fig. 4b data reveals a tripled unit cell for the states at $\bar{n}=\pm 8/3$. Red arrows in Fig. 4b and horizontal green bands in Fig. 4c mark features extending across the data independent of density. These so-called Brown–Zak oscillations[36] arise in Hofstadter systems whenever the magnetic field is tuned such that $b=p/q$ in Eq. (1) for mutually prime integers $p$ and $q$. In this case, the translational symmetry is restored and the effective quasiparticles move in zero effective magnetic field. Plotting $R_{xx}$ vs. $1/B$ shows a periodic oscillation in Fig. 4d. The longest period and largest amplitude oscillations are expected to occur when $p=1$[36] . The Fig. 4d inset shows the $N$th minimum (maximum) plotted vs. their number. The data closely follows a straight line from which the oscillation period in $1/B_F$ is extracted. Using $B_F A_e=h/e$ at the minima, where $A_e$ is the effective unit cell area, we find a unit cell area of $A_e=(1.94\pm0.03) \times 10^{-16}$ m$^2$. Based on the twist angle of 1.57º, we find the moiré unit cell area is $A=(6.8\pm0.27) \times 10^{-17}$ m$^2$. The ratio $A_e/A \approx 2.85\pm 0.1$ is consistent with a 3:1 reconstruction of the effective unit cell area. This reconstruction extends over a finite range in density. The observed tripled unit cell and the observed zero Chern number provide strong evidence for either armchair or star phases. The two phases are close in energy, with armchair being slightly favored, reflecting orbital geometric frustration. In addition, the four-fold degeneracy of LL at CNP indicates C3 breaking, compatible with armchair phase. Finally, we note that the evidence for the 3-fold unit cell expansion is not as clear when crossing to higher density than 8/3 filling (Fig. 4e). This may reflect a change in the unit cell that accompanies the reset behavior and Berry phase change that as discussed above.

In conclusion, we find at larger-than-magic twist angles, CI features at 1/3-fillings dominate over correlated insulating features integer $\bar{n}$ fillings, arising from repulsive interactions among orbitals extended over several unit cells. By varying the twist angle, we are able to tune the ground state from an "arm-chair" phase to a "brick wall" phase. In the latter, the geometric frustration is expected to lead to a large entropy, which may be measured in future experiments and would signal a novel example of geometric frustration originating from the fragile topology of the twisted bilayer band structure.

**Acknowledgement**
We thank Fan Zhang for helpful discussions. The experiments are supported by DOE BES Division under grant no. DE-SC0020187. Devices are fabricated using the nanofabrication facility that is supported by NSF Materials Research Science and Engineering Center Grant DMR-2011876. D.M. and E-A.K. are supported by the Gordon and Betty Moore Foundation's EPiQS Initiative, Grant GBMF10436. K.Z. is supported by the NSF under EAGER OSP-136036 and the Natural Sciences and Engineering Research Council of Canada (NSERC) under PGS-D-557580-2021. E-A.K. acknowledges support by the MURI grant FA9550-21-1-0429, the Ewha Frontier 10-10 Research Grant, and the Simons Fellowship in Theoretical Physics award 920665. A portion of this work was performed at the National High Magnetic Field Laboratory, which is supported by the National Science Foundation through NSF/DMR-1644779 and the State of Florida. Growth of hexagonal boron nitride crystals was supported by the Elemental Strategy Initiative conducted by the MEXT, Japan (Grant Number JPMXP0112101001) and JSPS KAKENHI (Grant Numbers 19H05790, 20H00354 and 21H05233).

**Fig. 1.** Temperature-dependent data from device D1 with $\theta=1.32°$. (a). Two-terminal $G$ vs. $V_g$ (bottom axis) and $\bar{n}$ at T=0.4, 0.6, 0.7, 0.95, 2, 5, 10, 15, 20, 25.6, 33.5, 37.6, 45, 62.6, and 72.3K, respectively (blue to purple). The grey and cyan vertical bands outline the features at integer and fractional fillings of the moiré superlattice, respectively. (b). Arrenhius plot for the conductance valley at $\bar{n}=\pm 8/3$. The lines are fits to a double exponential function. (c). Arrenhius plot at $\bar{n}=\pm 4/3$ (adjacent data points are connected by straight lines).

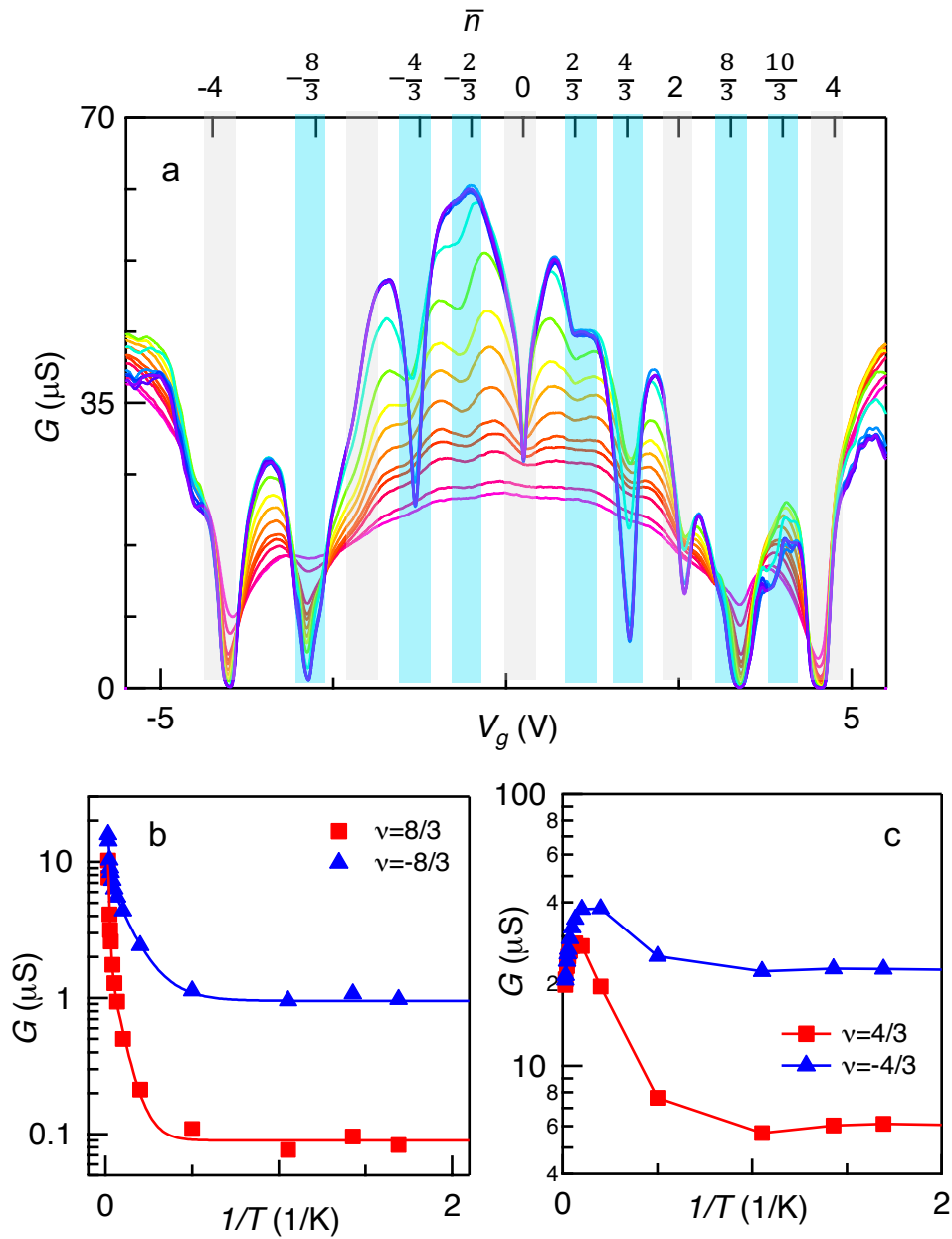

**Fig 2.** (a) Wannier orbitals at BA sites for two different twist angles. The insets denote the difference between the WOs of the two valleys. (b) Model with direct density-density interaction (V terms) and direct exchange (J terms) up to third nearest neighbor. (c) Schematic of states close in energy at $\bar{n}$=-8/3. The charge configurations at $\bar{n}$=-4/3 can be obtained from turning unoccupied sites in the configuration into doubly occupied sites (see SI). (d) Schematic of the WOs (left) and the competition (right) between the spin singlet, valley polarized state at zero field and the spin polarized, valley anti-aligned state at finite field for the sites with double occupation at $\bar{n}$=±4/3.

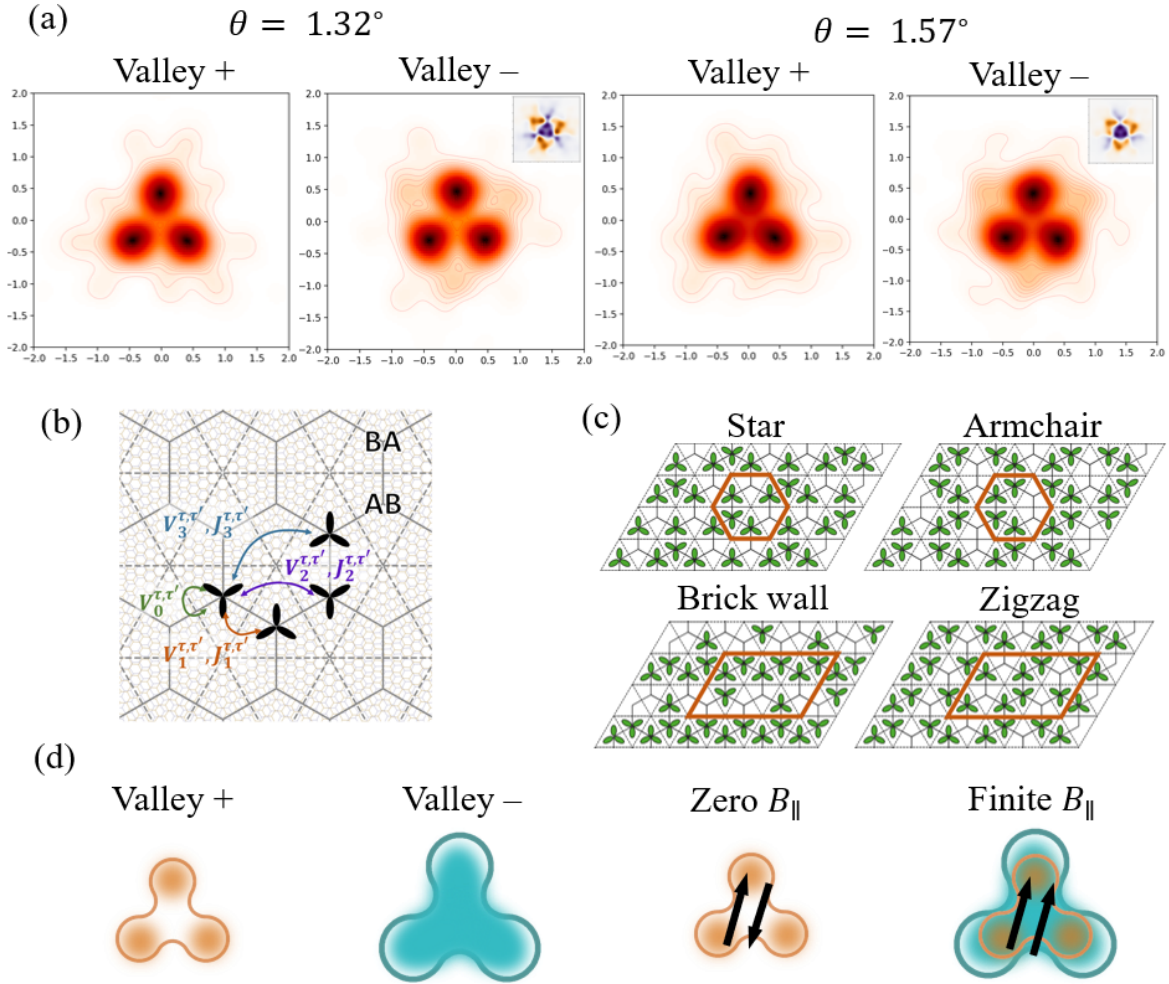

**Fig. 3.** Magnetotransport data of device D1 at *T*=300 mK. (a-b). Two-terminal $G(\bar{n}, B)$ in µS and Wannier diagram identifying features in a. Numbers indicate corresponding *t* quantum numbers. (c-d). $G(B)$ at $\bar{n}=\pm 4/3$ and $\pm 8/3$, respectively. (e-f). $G(\bar{n}, B_{||})$ in µS and line traces at $B_{||}$=0T, 3T, 6T, 12T and 18T (top to bottom). (g-h). $G(B_{||})$ at $\bar{n}=\pm 4/3$ and $\pm 8/3$, respectively.

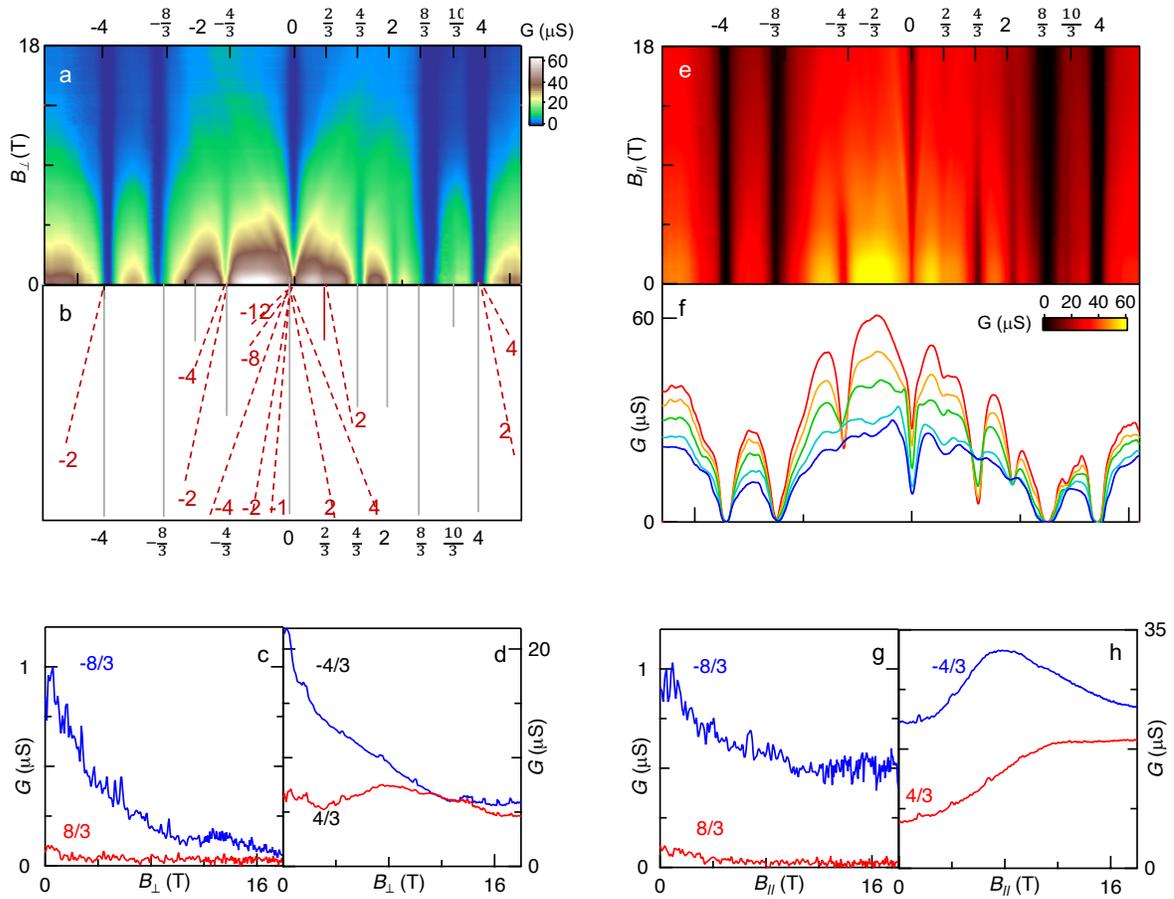

Fig. 4. Magnetotransport data of device D2 at $T$=1.5 K. (a-b). Four-terminal longitudinal resistance $R(\bar{n})$ at $B$=0 and $R((\bar{n}, B)$ in kΩ. (c). Wannier diagram identifying features in b. Numbers indicate corresponding $t$ quantum numbers. Horizontal green bands mark Brown-Zak oscillations. (d) Line trace taken at filling 8/3 vs. 1/$B$ (a polynomial background fit was subtracted). Inset, 1/$B$ values of $N$th maxima ($N$th-1/2 minima) extracted from main panel. The slope of the fitted line enables a determination of the period in 1/$B$. (e) Zoom-in plot of region in (b) near $\bar{n}$=8/3, showing the density-independent resistance minima in Brown-Zak bands.

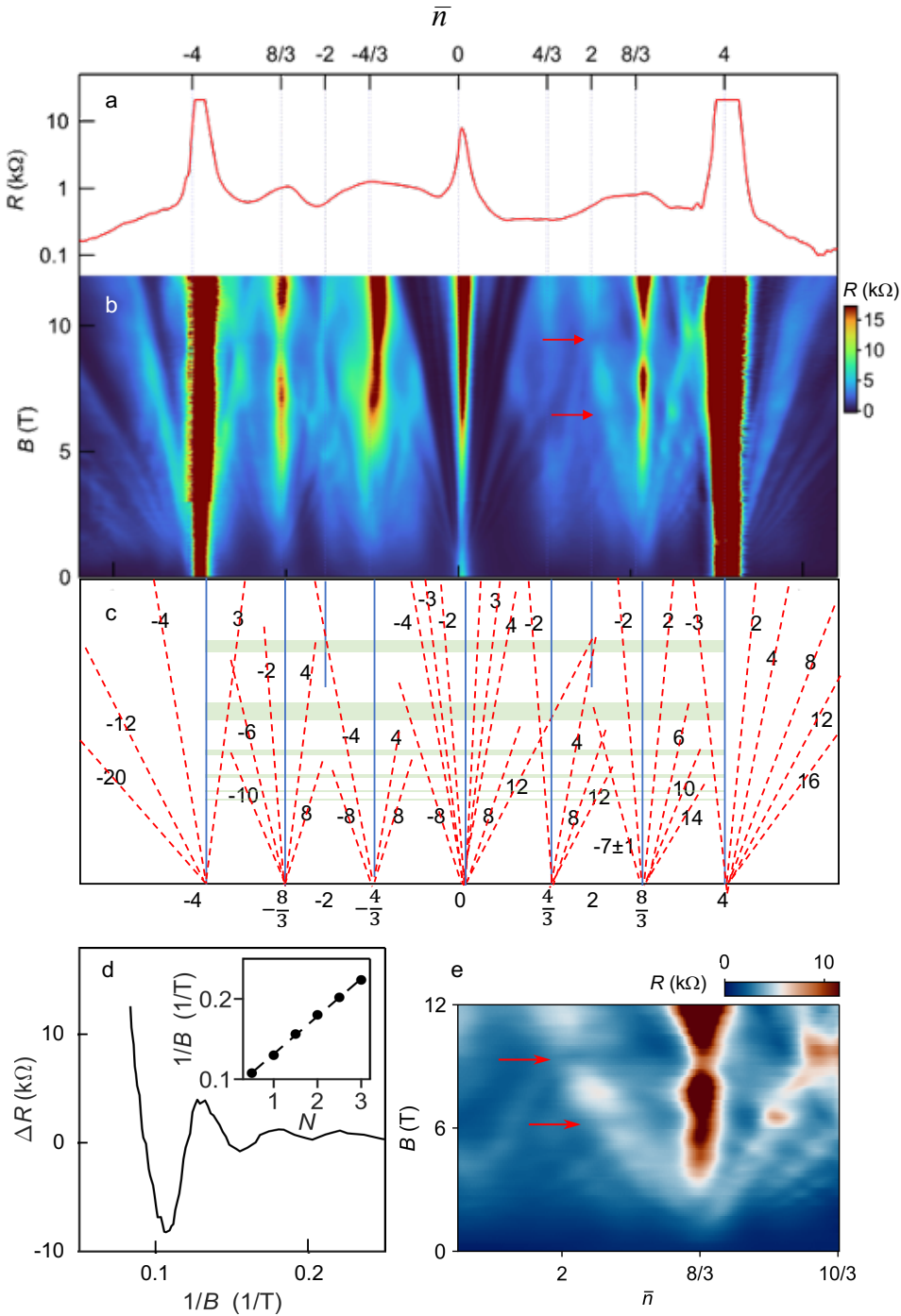